\documentclass[twocolumn,showpacs,prl,amsmath,amssymb,nofootinbib,nobalancelastpage]{revtex4}
\usepackage{psfig}
\usepackage[mathscr]{eucal}

\DeclareSymbolFont{EUsymb}{U}{eur}{m}{n}
\DeclareMathSymbol{\varmu} {\mathcal}{EUsymb}{"16}
\newcommand{\<}[1]{\langle {#1} \rangle}
\newcommand{\vhat}[1]{\hat{{\bf #1}}}
\newcommand{\tinyplus}{\mbox{\tiny\ensuremath{+}}}
\newcommand{\tinyminus}{\mbox{\tiny\ensuremath{-}}}
\newcommand{\tinyplusminus}{\mbox{\tiny\ensuremath{+\hspace{-0.03cm}-}}}

\newcommand{\tinyD}{\mbox{\tiny D}}
\newcommand{\tinyH}{\mbox{\tiny H}}
\newcommand{\tinyX}{\mbox{\tiny X}}
\newcommand{\tinyXH}{\mbox{\tiny XH}}
\newcommand{\tinyDH}{\mbox{\tiny DH}}
\newcommand{\tinyHH}{\mbox{\tiny HH}}
\newcommand{\tinyMuH}{\mbox{\tiny \ensuremath{\mu}H}}
\newcommand{\tinyB}{\mbox{\tiny B}}
\newcommand{\tinyN}{\mbox{\tiny N}}
\newcommand{\sigHH}{\sigma_{\tinyplusminus}^{\tinyHH}} 
\newcommand{\sigDH}{\sigma_{\tinyplusminus}^{\tinyDH}} 
\newcommand{\barsigHH}{\bar{\sigma}_{\tinyplusminus}^{\tinyHH}} 
\newcommand{\barsigDH}{\bar{\sigma}_{\tinyplusminus}^{\tinyDH}} 
\newcommand{\barsigXH}{\bar{\sigma}_{\tinyplusminus}^{\tinyXH}} 
\newcommand{\barsigMuH}{\bar{\sigma}_{\tinyplusminus}^{\tinyMuH}}

\begin{document}

\title{Measuring the Primordial Deuterium Abundance During the Cosmic Dark Ages}
\author{Kris Sigurdson}
\email{ksigurds@tapir.caltech.edu}
\affiliation{California Institute of Technology, Mail Code 130-33, Pasadena, CA
91125}
\author{Steven R. Furlanetto}
\email{sfurlane@tapir.caltech.edu}
\affiliation{California Institute of Technology, Mail Code 130-33, Pasadena, CA
91125}


\begin{abstract}

We discuss how measurements of fluctuations in the absorption of cosmic microwave background (CMB) photons by neutral gas during the cosmic dark ages, at redshifts $z \approx 7$--$200$, could reveal the primordial deuterium abundance of the Universe.  The strength of the cross-correlation of brightness-temperature fluctuations due to resonant absorption of CMB photons in the 21-cm line of neutral hydrogen with those due to resonant absorption of CMB photons in the 92-cm line of neutral deuterium is proportional to the fossil deuterium to hydrogen ratio [{\rm D/H}] fixed during big bang nucleosynthesis (BBN).  Although technically challenging, this measurement could provide the cleanest possible determination of [\rm D/H], free from contamination by structure formation processes at lower redshifts, and has the potential to improve BBN constraints to the baryon density of the Universe $\Omega_{\rm b} h^2$.  We also present our results for the thermal spin-change cross-section for deuterium-hydrogen scattering, which may be useful in a more general context than we describe here.

\end{abstract}


\pacs{98.80.-k,98.80.Ft,32.10.Fn,34.50.-s,95.30.Dr,95.30.Jx,98.70.Vc}

\maketitle
\noindent \emph{Introduction ---} After the cosmic microwave background (CMB) radiation decoupled from the baryons at a redshift \mbox{$z \approx 1100$}, most CMB photons propagated unfettered through the neutral primordial medium.  This has allowed exquisite measurements of the temperature fluctuations in the primordial plasma at the surface of last scattering, and the statistical properties of these fluctuations have recently been used, in conjunction with other observations, to determine the cosmology of our Universe \cite{Spergel:2003cb}. After the photons kinetically decoupled from the gas at $z \sim 200$, the latter cooled adiabatically with $T_{\rm g} \propto (1+z)^{2}$, faster than the $T_{\gamma} \propto (1+z)$ cooling of the CMB.  This epoch, with most of the baryons in the form of relatively cold neutral atoms and before the first stars formed, is known as the cosmic dark ages.

The reason most CMB photons propagate unimpeded through the neutral primordial gas is elementary quantum mechanics --- atoms absorb non-ionizing radiation only at the discrete wavelengths determined by the differences of their atomic energy levels.  One interesting example is the well-known 21-cm spin-flip transition \cite{FermiHFS}, due to the hyperfine splitting of the ground state of the hydrogen (H) atom.  At any given $z$, CMB photons with wavelength $\lambda_{21}=21.1$~cm can resonantly excite this transition.  By measuring brightness-temperature fluctuations due to density fluctuations in the neutral gas \cite{HoganScottRees}, radio telescopes observing at $\lambda = (1+z)\lambda_{21}$ can probe the matter power spectrum at $z \approx 30$--$200$ \cite{Loeb:2003ya}.

In this \emph{Letter} we discuss another application of these measurements.  Less well-known than the 21-cm transition of neutral H is the spin-flip transition of neutral deuterium (D) at $\lambda_{92}=91.6$~cm \cite{Nafe1947,DHmeasure}.  We show below that cross-correlating brightness-temperature fluctuations at a wavelength $\lambda_{\tinyH} = (1+z)\lambda_{21}$ with those at a wavelength $\lambda_{\tinyD} = (1+z)\lambda_{92}$ allows a measurement of the primordial D abundance.  In principle, this technique could constrain the primordial value of $[{\rm D/H}]\equiv n_{\tinyD}/n_{\tinyH}$ to better than $1\%$.  While there is no physical obstacle to such a measurement, it would certainly be technically challenging, and require a heroic experimental effort;  simply detecting neutral D during the cosmic dark ages would be a significantly easier goal. 

Deuterium has long been recognized as our best `baryometer' because its primeval relic abundance is so sensitive to the baryon-to-photon ratio $\eta=n_{\rm b}/s$.  Moreover, big bang nucleosynthesis (BBN) \cite{Wagoner:1966pv} is the only known natural production mechanism, although mechanisms inside galaxies can destroy it  \cite{Reeves1973}. The measurement we describe below could thus determine the true BBN abundance of D and, in principle,  might improve BBN constraints to the baryon density of the Universe $\Omega_{\rm b} h^2$.

\noindent \emph{Hyperfine Structure of H and D Atoms ---} The $\pmb{\varmu}\cdot\mathbf{B}$ interaction between the magnetic moments of the electron and the nucleus splits the ground state of single-electron atoms into eigenstates of the total spin operator $\mathbf{F}=\mathbf{S}+\mathbf{I}$ with eigenvalues $F_{\tinyplus}=I+1/2$ and $F_{\tinyminus}=I-1/2$ and $\Delta E = (16/3)F_{\tinyplus}\varmu_{\tinyB}(g_{\tinyN}\varmu_{\tinyN}/a_{0}^3)$ (e.g., \cite{QMref}).  Here, $\mathbf{S}$ is electron spin, $\mathbf{I}$ is nuclear spin, $a_{0}$ is the Bohr radius, $\varmu_{\tinyB}$ is the Bohr magneton, $\varmu_{\tinyN}$ is the nuclear magneton, and $g_{\tinyN}$ is the nuclear $g$ factor ($g_{\rm p}=5.56$ for H; $g_{\tinyD}=0.857$ for D).  The proton, with $I=1/2$, splits the H ground state into a triplet with $F_{\tinyplus}=1$ and a singlet with $F_{\tinyminus}=0$.  The deuteron, with $I=1$, splits the D ground state into a quartet with $F_{\tinyplus}=3/2$ and a doublet with $F_{\tinyminus}=1/2$.  

\noindent \emph{The Spin-Temperature ---} The population of atoms in the excited spin state relative to the ground state $n_{\tinyplus}/n_{\tinyminus}=(g_{\tinyplus}/g_{\tinyminus}){\rm exp}\{-T_{\star}/T_{\rm s}\}$ can be characterized by a spin temperature $T_{\rm s}$.  Here $g_{\tinyplus}=2F_{\tinyplus}+1$ and $g_{\tinyminus}=2F_{\tinyminus}+1$ are the spin degeneracy factors and $T_{\star}=\Delta E/k_{\tinyB}$.  For H and D, we have respectively $T_{\star}^{\tinyH}=0.0682$~K, $T_{\star}^{\tinyD}=0.0157$~K, and $(g_{\tinyplus}^{\tinyH}/g_{\tinyminus}^{\tinyH})=3$, $(g_{\tinyplus}^{\tinyD}/g_{\tinyminus}^{\tinyD})=2$.

Competition between three factors determines $T_{\rm s}$: absorption of 21-cm CMB photons, absorption and re-emission of Lyman-$\alpha$ photons (the Wouthuysen-Field  or WF effect \cite{WouthuysenField,Field58}), and atomic spin-change collisions (collisions with free electrons are unimportant in these environments \cite{Field58}). The first drives $T_{\rm s}$ toward $T_{\gamma}$, while the latter two drive it toward the gas temperature $T_{\rm g}$.  In equilibrium the spin temperature of the a species X (either D or H) is
\begin{align}
T_{\rm s}^{\tinyX}=\frac{(1+\chi^{\tinyX})T_{\rm g}T_{\gamma}}{(T_{\rm g}+\chi^{\tinyX} T_{\gamma})} \, ,
\end{align}
where $\chi^{\tinyX} \equiv \chi_{c}^{\tinyX} + \chi_{\alpha}^{\tinyX}$ is the sum of the equilibrium threshold parameters for spin-change collisions and for radiative coupling through the WF effect.  Explicitly, $\chi_{c}^{\tinyX}=(C_{\tinyplusminus}^{\tinyX}T_{\star}^{\tinyX})/(A_{\tinyplusminus}^{\tinyX}T_{\gamma})$ and $\chi_{\alpha}^{\tinyX}=(P_{\tinyplusminus}^{\tinyX}T_{\star}^{\tinyX})/(A_{\tinyplusminus}^{\tinyX}T_{\gamma})$, where $C_{\tinyplusminus}^{\tinyX}$ is the collisional de-excitation rate, $A_{\tinyplusminus}^{\tinyX}$ is an Einstein coefficient, and $P_{\tinyplusminus}^{\tinyX} \propto P_{\alpha}$, where $P_{\alpha}$ is the total Lyman-$\alpha$ scattering rate.  At $z \gg 10$, before the first galaxies formed, $P_{\alpha}$ is tiny and the WF effect can be neglected.  However, it might have interesting consequences near $z \sim 10$.

\noindent \emph{H-H and D-H Collision Rates ---}   While the cross section for H-H spin-change collisions $\sigHH$ is well known \cite{Dalgarno:1961,Smith:1966,Allison:1969,Allison:1972,Zygelman:2005}, we were unable to locate the D-H spin-change cross section for the temperature range of interest and computed $\sigDH$ using standard methods\footnote{It has been computed at higher temperatures by Ref.~\cite{Krstic:1999}, at 1~K by Ref.~\cite{Reynolds:1989}, and measured at 1~K by Ref.~\cite{Hayden:1995}.}.  In the elastic approximation, 
\begin{align} 
\sigDH=\frac{\pi}{3k^2}\sum_{l=0}^{\infty}(2l+1){\rm sin}^2({}^{t}\!\eta_{l}^{\tinyDH}-{}^{s}\!\eta_{l}^{\tinyDH}) \, ,
\end{align}
where $k=\mu_{\tinyDH} v/\hbar$, $\mu_{\tinyDH}=m_{\tinyD}m_{\tinyH}/(m_{\tinyD}+m_{\tinyH})$ is the reduced mass of the D-H system, and $v$ is the relative velocity.\footnote{Compare with $\sigHH=(\pi/4k^2)\sum_{l=0}^{\infty}(2l+1){\rm sin}^2({}^{t}\!\eta_{l}^{\tinyHH}-{}^{s}\!\eta_{l}^{\tinyHH})$ for H-H collisions when quantum symmetry can be neglected \cite{Zygelman:2005}.}  The partial wave phase shifts in the triplet and singlet electronic potentials $V_{t}(R)$ and $V_{\rm s}(R)$ in which the D and H atoms scatter are ${}^{t}\!\eta_{l}^{\tinyDH}$ and ${}^{s}\!\eta_{l}^{\tinyDH}$ respectively.     We used the variational potentials of Refs.~\cite{Kolos:1986,Frye:1989} for $R \leq 12\,a_{0}$, the smooth fit of Ref.~\cite{Jamieson:1992} for $V_{t}(R)$ between $12\,a_{0} \leq R \leq 15\,a_{0}$, and the asymptotic form of Ref.~\cite{Kolos:1967} for larger $R$.   We found the phase shifts by solving the radial Schr\"odinger equation at each energy $E=\hbar^2 k^2/2\mu_{\tinyDH}$ and angular momentum $\sqrt{l(l+1)}\hbar$, truncating at $l_{max} \sim 100$ to resolve the resonant structure at sufficiently high $E$. We verified that our results agree with Ref.~\cite{Reynolds:1989} at 1~K (after accounting for our more recent potentials), with those of Refs.~\cite{Allison:1969,Allison:1972,Zygelman:2005,Shizgal:1979}\footnote{We are in harmony with Ref.~\cite{Zygelman:2005} which found that $\kappa(1 \rightarrow 0 )=\bar{v}_{\tinyHH}\barsigHH$ is 4/3 larger than previously quoted \cite{Smith:1966,Allison:1969} at high $T_{\rm g}$.}  for $\sigHH$, and with Ref.~\cite{Shizgal:1979} for the muonium-H spin-change cross section $\sigma_{\tinyplusminus}^{\tinyMuH}$.

\begin{figure}
\centerline{\psfig{file=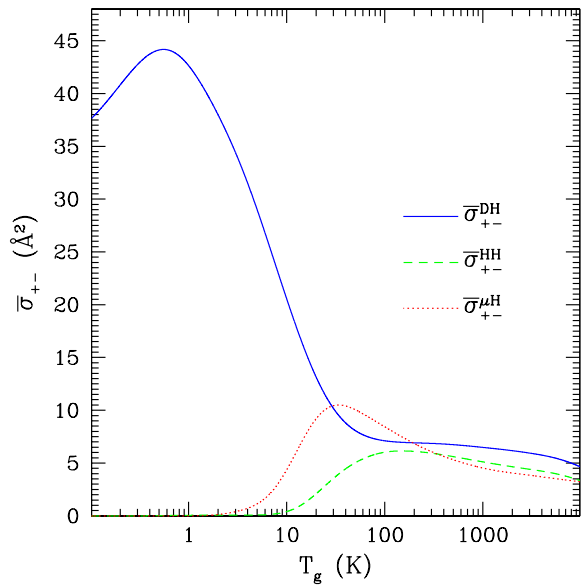,width=3.015in,angle=0}}
\caption{The thermal spin-change cross sections that keep $T_{\rm s}$ collisionally coupled to $T_g$ for D (solid line), H (dashed line), and muonium (dotted line).  Although the potentials are identical, the peaks differ because of the reduced masses.}
\label{fig:thermcross}
\end{figure} 

The collisional de-excitation rate is $C_{\tinyplusminus}^{\tinyX} = \bar{v}_{\tinyXH}\barsigXH n_{\tinyH}$, where $\bar{v}_{\tinyXH}=\sqrt{8 k_{\tinyB} T_{\rm g}/\mu_{\tinyXH}\pi}$ is the thermal velocity, $\barsigXH$ is the thermal spin-change cross-section (averaged over the Maxwell-Boltzmann distribution of relative velocities), and $n_{\tinyH}$ is the number density of H atoms. 
In Fig.~{\ref{fig:thermcross} we plot $\barsigHH$, $\barsigDH$, and $\barsigMuH$.  While $\barsigHH$ falls off for $T_{\rm g} \lesssim 100$~K, $\barsigDH$ continues to rise to a peak near $T_{\rm g} \sim 1$~K.  This occurs because of low-energy  $s$-wave and $p$-wave contributions to D-H scattering (a much larger scattering length and a scattering resonance).  These do not appear for H-H because of the differing reduced mass ($\mu_{\tinyHH} \approx m_{\tinyH}/2$ while $\mu_{\tinyDH} \approx 2m_{\tinyH}/3$).  The discussion of D-H spin-change in Ref.~\cite{Field58} did not account for this and incorrectly concluded that $\sigDH \sim \sigHH$.

\noindent \emph{Spin-Temperature Evolution ---} In Fig.~{\ref{fig:temp} we plot $T_\gamma$, $T_g$ (found using {\tt recfast} \cite{SeagerSasselovScott}), $T_{\rm s}^{\tinyH}$, and $T_{\rm s}^{\tinyD}$ as a function of $z$.  After the gas cools below $T_\gamma$, collisions keep $T_{\rm s}^{\tinyH}$ and $T_{\rm s}^{\tinyD}$ coupled to $T_{\rm g}$.  Near $z \sim 30$ collisions become inefficient for H and $T_{\rm s}^{\tinyH}$ returns to $T_{\gamma}$. $T_{\rm s}^{\tinyD}$ remains coupled to $T_{\rm g}$ down to significantly lower redshift both because the lifetime of the excited state of D is relatively long ($A_{\tinyplusminus}^{\tinyH}/A_{\tinyplusminus}^{\tinyD}=61.35$) and because $\barsigDH \gg \barsigHH$ at low temperatures.
\begin{figure}
\centerline{\psfig{file=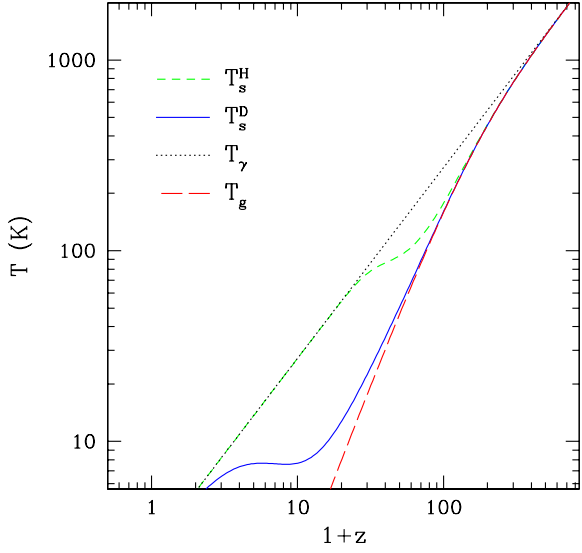,width=3.015in,angle=0}}
\caption{The H and D spin temperatures as a function of $z$.  
Here we assume $P_\alpha=0$ for all $z$.}
\label{fig:temp}
\end{figure}

\noindent \emph{Brightness Temperature Fluctuations ---} When the spin temperature of a given species is less than $T_{\gamma}$ it will absorb CMB photons.  The brightness temperature is $T_{\rm b}^{\tinyX}=a \tau_{\tinyX}(T_{\rm s}^{\tinyX}-T_{\gamma})$, where 
\begin{align}
\tau_{\tinyX}=\frac{g_{\tinyplus}^{\tinyX}c\lambda^2 h A_{\tinyplusminus}^{\tinyX}n_{\tinyX}}{8 (g_{\tinyplus}^{\tinyX}+g_{\tinyminus}^{\tinyX})\pi k_{\tinyB} T_{\rm s}^{\tinyX} {\cal H}(z)}
\end{align}
is the optical depth of the spin-flip transition in question, $a=1/(1+z)$ and ${\cal H}(z)$ is the Hubble parameter. 
We are interested in correlations between brightness temperature fluctuations $\delta_{T_{\rm b}}^{\tinyX}(\vhat{n},a) \equiv \delta T_{\rm b}^{\tinyX}(\vhat{n},a)/T_{\rm b}^{\tinyX}(a)=\beta_{T_{\rm b}}^{\tinyX}(a)\delta(\vhat{n},a)$ observed in a direction $\vhat{n}$ at wavelengths differing by a factor $\lambda_{92}/\lambda_{21}$.  Here,
\begin{align}
\beta_{T_{\rm b}}^{\tinyX} = 1+\frac{\chi_{\rm c}^{\tinyX}}{\widehat{\chi}^{\tinyX}}+ \Gamma\left[\frac{T_{\gamma}}{T_{\rm g}-T_{\gamma}} + \frac{\chi_{\rm c}^{\tinyX}}{\widehat{\chi}^{\tinyX}} \frac{d {\rm ln}(C_{\tinyplusminus}^{\tinyX})}{d {\rm ln}(T_{\rm g})}\right] \, , 
\label{eqn:fluctuations}
\end{align}
$\delta(\vhat{n},a)=\delta n_{\tinyH}(\vhat{n},a)/n_{\tinyH}(a)=\delta n_{\tinyD}(\vhat{n},a)/n_{\tinyD}(a)$ is the density contrast, and $\widehat{\chi}^{\tinyX} \equiv \chi^{\tinyX}(1+\chi^{\tinyX})$.  At high $z$, when $T_{\rm g} \approx T_{\gamma}$,  $\Gamma \rightarrow 0$ due to residual Thomson scattering with free electrons \cite{PeeblesCosmology} (fluctuations are isothermal), but as the gas begins to cool adiabatically $\Gamma \rightarrow 2/3$ \cite{Bharadwaj:2004nr}.  In Eq.~(\ref{eqn:fluctuations}) we have neglected the contributions to $\delta_{T_b}$ from fluctuations in the neutral fraction (likely to be small at high $z$) and, for simplicity, fluctuations in the gradient of the radial velocity $\delta_{\partial_r v_r}$ \cite{Barkana:2004zy}.  The latter will enhance our signal by a factor of $\sim 1$--$2$.  In Fig.~{\ref{fig:deltatb} we plot $T_{\rm b}^{\tinyH}$, $a\beta_{T_{\rm b}}^{\tinyH}T_{\rm b}^{\tinyH}$, $\widetilde{T}_{\rm b}^{\tinyD}$ and $a\beta_{T_{\rm b}}^{\tinyD}\widetilde{T}_{\rm b}^{\tiny d}$, where $T_{\rm b}^{\tinyD}\equiv \epsilon \widetilde{T}_{\rm b}^{\tinyD}$ and $\epsilon \equiv [{\rm D/H}]$.  We see that $\widetilde{T}_{\rm b}^{\tinyD}$ and $a\beta_{T_{\rm b}}^{\tinyD}\widetilde{T}_{\rm b}^{\tiny d}$ peak at much lower $z$ than their H counterparts because, as discussed above, $T_{\rm s}^{\tinyD}$ is coupled to $T_{\rm g}$ to lower $z$.
 \begin{figure}[t]
\centerline{\psfig{file=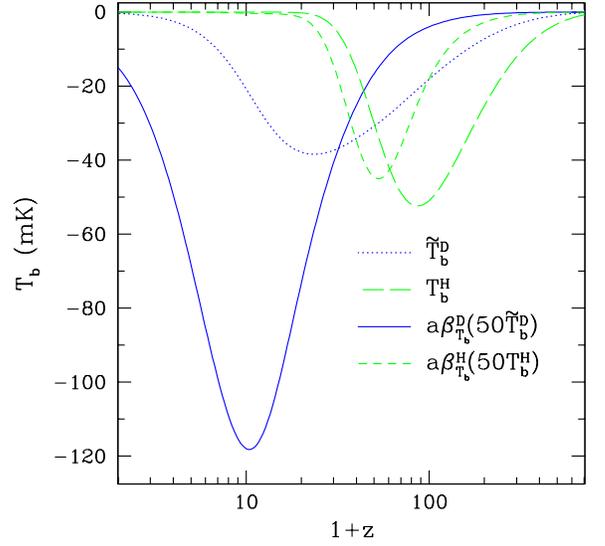,width=3.015in,angle=0}}
\caption{The brightness temperatures $T_{\rm b}^{\tinyH}$ and $\widetilde{T}_{\rm b}^{\tinyD}$ and the brightness temperature fluctuations $\beta_{T_{\rm b}}^{\tinyH}T_{\rm b}^{\tinyH}$ and $\beta_{T_{\rm b}}^{\tinyD}\widetilde{T}_{\rm b}^{\tinyD}$ (scaled to the growth rate of density perturbations $\delta \propto a$).}
\label{fig:deltatb}
\end{figure}

\noindent \emph{D-H Cross Correlations ---} We now estimate the cross-correlation of brightness temperature fluctuations across frequencies related by $\lambda_{21}/\lambda_{92}$.  We write the brightness temperature fluctuation due to H or D as $H(\vhat{n},a)=\beta_{ T_{\rm b}}^{\tinyH}(a)T_{\rm b}^{\tinyH}(a)\delta(\vhat{n},a)$ and $\epsilon D(\vhat{n},a)= \epsilon \beta_{ T_{\rm b}}^{\tinyD}(a)\widetilde{T}_{\rm b}^{\tinyD}(a)\delta(\vhat{n},a)$ respectively.  A radio telescope observing at a frequency $\nu$ will measure the quantity ${\cal O}[\vhat{n};\nu]=H(\vhat{n},\nu/\nu_{21})+\epsilon D(\vhat{n},\nu/\nu_{92})+N[\vhat{n};\nu]$, where $N[\vhat{n};\nu]$ is the instrumental noise.  We form the product ${\cal O}[\vhat{n};\nu_{\alpha}]{\cal O}[\vhat{n};\nu_{\beta}]$, where $\nu_{\beta} \equiv (\nu_{92}/\nu_{21})\nu_{\alpha}$.  Assuming that $\delta(\vhat{n},a)$ is a zero-mean Gaussian random field and uncorrelated Gaussian noise, its expectation value is $\<{{\cal O}[\vhat{n};\nu_{\alpha}]{\cal O}[\vhat{n};\nu_{\beta}]}=\epsilon\<{H_\alpha D_\beta}$ to leading order in $\epsilon$.  Here we have introduced the shorthand $H_{\alpha}\equiv H(\vhat{n},\nu_{\alpha}/\nu_{21})$, $D_{\beta} \equiv D(\vhat{n},\nu_{\beta}/\nu_{92})$, and $N_{\alpha}=N[\vhat{n};\nu_{\alpha}]$.  

We now understand the crucial point of this \emph{Letter}.  \emph{The 21-cm and 92-cm fluctuations at these frequency separations must be correlated because they trace the same underlying patches of the Universe.}

Note that we have neglected the relatively small intrinsic correlations of the H brightness fluctuations from large-scale modes of the density field which contribute at the level of $\sim\!0.1$\% or less of the D signal.
If necessary, these intrinsic H correlations could be removed through independent measurements of the matter power spectrum or by correlating the 21-cm signal with frequencies near but not equal to the corresponding D patches.

\noindent \emph{Signal Estimate ---}
The signal-to-noise contributed by a pair of frequency bands centered around $(\nu_{\alpha},\nu_{\beta})$ for an experiment with a maximum baseline of $L$ and frequency resolution $\Delta \nu$ is
\begin{align}
\frac{{\cal S}}{{\cal N}}(\nu_\alpha)=\epsilon \frac{4}{\theta_{\beta}}\frac{\<{H_{\alpha}D_{\beta}}}{\sqrt{(\<{H_{\alpha}^2}+\<{N_{\alpha}^2})(\<{H_{\beta}^2}+\<{N_{\beta}^2}}}
\label{eq:sn}
\end{align}
where $\theta_\beta=\lambda_{\beta}/L$ is the angular resolution in the D band.  Here, $\<{H_{\alpha}D_{\beta}}=\sigma_{\delta}^2(\beta_{T_{\rm b}}^{\tinyH}T_{\rm b}^{\tinyH})(\beta_{T_b}^{\tinyD}\widetilde{T}_{\rm b}^{\tinyD})$ and $\<{H_{\alpha}^2} = \sigma_{\delta}^2(\beta_{T_{\rm b}}^{\tinyH}T_{\rm b}^{\tinyH})^2$, where 
\begin{align}
\sigma_{\delta}^2=\frac{2}{\pi^2}\int_{0}^{\infty}dk kP(k)\int_{0}^{k}dk_{z} j_0^2(\xi k_{z} \rho) \frac{J_{1}^2(\sqrt{k^2-k_z^2}\rho)}{(k^2-k_z^2)\rho^2}
\end{align}
is the variance in the density field at scale-factor $a$ smoothed over the coin-shaped regions of the Universe of comoving radius $\rho$ and thickness $2\xi \rho$ that are sampled by an experiment with high spectral (radial) resolution but lower angular (transverse) resolution.  We adopt a noise variance of $\<{N_{\alpha}^2}= T_{\rm sys}^2/(f_{\rm cov}^2\, \Delta\nu\, t_{\rm int})$ (e.g., \cite{Zald04}), where $T_{\rm sys}=6500 [\nu_{\alpha}/(30{\rm MHz})]
^{-2}~{\rm K}$ is the noise temperature, $f_{\rm cov}$ is the covering fraction of the array, and $t_{\rm int}$ is the integration time.  
Our choice for $T_{\rm sys}$ is only an estimate, and the noise (ultimately due to Galactic synchrotron radiation) varies strongly across the sky.  The total signal-to-noise ratio $({\cal S/N})_{\rm tot}$ is the sum in quadrature over all pairs of frequency bands $(\nu_{\alpha},\nu_{\beta})$. 

If collisions dominate the coupling between $T_{\rm s}^{\tinyH}$ and $T_{\rm g}$ down to $z \sim 7$ then a value $[{\rm D/H}] \sim 3 \times 10^{-5}$ could be detected at 1- to 2-$\sigma$ by an experiment with $L \sim 7.5$~km and $\Delta \nu \sim 100$~kHz in $\sim$6 years.   If, however, the first generation of stars created a flux of Lyman-$\alpha$ photons which coupled $T_{\rm s}^{\tinyH}$ to $T_{\rm g}$ until $z \sim 7$ through the WF effect
without significantly heating the gas, a similar detection might be made by a smaller experiment with $L \sim 2.5$~km and $\Delta \nu \sim 1$~kHz.  
Although it dramatically enhances the 21-cm signal, the WF effect does not improve the [{\rm D/H}] measurement by the same margin because ${\cal S/N}$ becomes independent of $H_\alpha$ once $\<{H_\alpha^2} \gg \<{N_\alpha^2}$ (it only serves to make the H fluctuations a better matched template). Finally, we note that an experiment capable of mapping $21$-cm brightness-temperature fluctuations out to $l_{max} \sim 10^{5}$ (where it may be a powerful probe of the small-scale matter power spectrum \cite{Loeb:2003ya}) could measure [{\rm D/H}] to a precision as good as $ \sim 1$\% --- or even $\sim 0.1\%$ if the WF effect coupling is efficient.

 For these estimates we have assumed a $\Lambda$CDM cosmology with $n_{\rm s}=1$, and that a significant fraction of the Universe remains neutral until $z \sim 7$.  The largest contribution to the signal originates from $z \lesssim 10$ where the D signal peaks and the variance in the density fluctuations is largest.  Varying these assumptions, or including additional sources of temperature fluctuations,  could change $({\cal S/N})_{\rm tot}$ by factors of order unity.

\noindent \emph{Discussion ---}
Despite the obvious technical challenges in observing this signal, we emphasize that it has the virtue of providing the cleanest possible measurement of the primordial [{\rm D/H}], free from contamination by structure formation processes at lower $z$.  Via the window of BBN, this would allow radio telescopes to peer into the first few minutes of the Universe.  We believe future searches for cosmic $21$-cm fluctuations should bear this possibility in mind.

We also note that $^3$He$^+$ has a hyperfine transition (with $\lambda=3.46$~cm) that can be used in a similar fashion; it has the advantage of much lower foreground contamination at higher frequencies.  This line will appear during reionization and should exhibit a strong anti-correlation with the corresponding 21-cm signal.  If the astrophysics of reionization can be understood well enough, the cross-correlation of this line with the 21-cm line could supplement the D-H experiment in order to probe BBN in even more detail.

\noindent \emph{Acknowledgments ---} We thank A. Dalgarno, J. Gair, M. Hayden,
M. Kamionkowski, T. Pearson, M. Reynolds, D. Scott and B. Zygelman for discussions.  KS
acknowledges the support of a Canadian NSERC Postgraduate Scholarship.
This work was supported in part by NASA NAG5-9821 and DoE
DE-FG03-92-ER40701.


\begin{thebibliography}{1}

\bibitem{Spergel:2003cb}
D.~N.~Spergel {\it et al.},
Astrophys.\ J.\ Suppl.\  {\bf 148}, 175 (2003).

\bibitem{FermiHFS}
E.~Fermi, 
Zeits.\ f.\ Physik  {\bf 60}, 320 (1930).

\bibitem{HoganScottRees}
C.~J.~Hogan and M.~J.~Rees,
Mon.\ Not.\ Roy.\ Astron.\ Soc.\  {\bf 188}, 791 (1979);
D.~Scott and M.~J.~Rees,
Mon.\ Not.\ Roy.\ Astron.\ Soc.\  {\bf 247}, 510 (1990).

\bibitem{Loeb:2003ya}
A.~Loeb and M.~Zaldarriaga,
Phys.\ Rev.\ Lett.\  {\bf 92}, 211301 (2004).

\bibitem{Nafe1947}
J.~E.~Nafe, E.~B.~Nelson, and I.~I.~Rabi,
Phys.\ Rev.\  {\bf 71} 914 (1947).

\bibitem{DHmeasure}
S.~Weinreb,
Nature {\bf 195} 367 (1962);
J.~M.~Pasachoff and D.~A.~Cesarsky,
Astrophys.\ J.\  {\bf 193}, 65 (1974);
K.~R.~Anantharamaiah and V.~Radhakrishnan,
Astron.\ Astrophys. {\bf 79}, L9 (1979);
L.~Blitz and C.~Heiles, 
Astrophys.\ J.\ Lett.\ {\bf 313}, L95 (1987);
J.~N.~Chengalur, R.~Braun and W.~B.~Burton,
Astron.\ Astrophys. {\bf 318}, L35 (1997).

\bibitem{Wagoner:1966pv}
R.~V.~Wagoner, W.~A.~Fowler and F.~Hoyle,
Astrophys.\ J.\  {\bf 148}, 3 (1967).

\bibitem{Reeves1973}
H.~Reeves, J.~Audouze, W.~A.~Fowler, and D.~N.~Schramm,
Astrophys.\ J.\  {\bf 179}, 909 (1973).

\bibitem{QMref}
C.~Cohen-Tannoudji, B.~Diu, and F.~Lalo\"e,
{\it Quantum Mechanics, Vol. 2.} (Wiley, New York, 1977).

\bibitem{Field58}
G.~B.~Field, 
Proc.\ IRE {\bf 46}, 240 (1958).

\bibitem{WouthuysenField}
S.~A.~Wouthuysen,
Astron.\ J.\ {\bf 57}, 31 (1952).

\bibitem{Dalgarno:1961}
A.~Dalgarno,
Proc.\ R.\ Soc.\ London\ Ser.\ A {\bf 262}, 132 (1961).

\bibitem{Smith:1966}
F.~J.~Smith,
Planet.\ Space \ Sci. {\bf 14}, 929 (1966).

\bibitem{Allison:1969}
A.~C.~Allison, and A.~Dalgarno,
Astrophys.\ J.\  {\bf 158}, 423 (1969).

\bibitem{Allison:1972}
A.~C.~Allison,
Phys.\ Rev.\ A {\bf 5}, 2695 (1972).

\bibitem{Zygelman:2005}
B.~Zygelman,
Astrophys.\ J.\  {\bf 622}, 1356 (2005).

\bibitem{Shizgal:1979}
B.~Shizgal,
J.\ Phys.\ B: At.\ Mol.\ Phys. {\bf 12}, 3611 (1979).

\bibitem{Krstic:1999}
P.~S.~Krstic and D.~R.~Schultz, 
Phys.\ Rev.\ A {\bf 60}, 2118 (1999).

\bibitem{Reynolds:1989}
M.~W.~Reynolds, 
Ph.D. thesis, University of British Columbia, 1989.

\bibitem{Hayden:1995}
M.~E.~Hayden and W.~N.~Hardy,
J.\ Low.\ Temp.\ Phys.\  {\bf 99}, 787 (1995).

\bibitem{Kolos:1986}
W.~Kolos, K.~Szalewicz, and H.~J.~Monkhorst,
J.\ Chem.\ Phys. {\bf 84}, 3278 (1986).

\bibitem{Frye:1989}
D.~Frye, G.~C.~Lie, and E.~Clementi,
J.\ Chem.\ Phys. {\bf 91}, 2366 (1989).

\bibitem{Jamieson:1992}
M.~J.~Jamieson, A.~Dalgarno, and J.~N.~Yukich,
Phys.\ Rev.\ A {\bf 46}, 6956 (1992).

\bibitem{Kolos:1967}
W.~Kolos,
Int.\ J.\ Quantum Chem.\ {\bf 1}, 169 (1967).

\bibitem{SeagerSasselovScott}
S.~Seager, D.~D.~Sasselov and D.~Scott,
Astrophys.\ J.\  {\bf 523}, L1 (1999);
S.~Seager, D.~D.~Sasselov and D.~Scott,
Astrophys.\ J.\ Suppl.\  {\bf 128}, 407 (2000).

\bibitem{PeeblesCosmology}
P.~J.~E.~Peebles, {\it Principles of Physical Cosmology} (Princeton University Press, Princeton, 1993).

\bibitem{Bharadwaj:2004nr}
S.~Bharadwaj and S.~S.~Ali,
Mon.\ Not.\ Roy.\ Astron.\ Soc.\  {\bf 352}, 142 (2004).

\bibitem{Barkana:2004zy}
R.~Barkana and A.~Loeb,
arXiv:astro-ph/0409572.

\bibitem{Zald04}
M.~Zaldarriaga, S.~R.~Furlanetto and L.~Hernquist,
Astrophys.\ J.\  {\bf 608}, 622 (2004).

\end{thebibliography}
\end{document}